\documentstyle[12pt]{article}
\begin{document}
\def\beq{\begin{equation}}
\def\eeq{\end{equation}}
\def\bea{\begin{eqnarray}}
\def\eea{\end{eqnarray}}
\def\ve{\vert}
\def\vel{\left|}
\def\ver{\right|}
\def\nnb{\nonumber}
\def\ga{\left(}
\def\dr{\right)}
\def\aga{\left\{}
\def\adr{\right\}}
\def\rar{\rightarrow}
\def\nnb{\nonumber}
\def\la{\langle}
\def\ra{\rangle}
\def\ba{\begin{array}}
\def\ea{\end{array}}
\def\tep{$B \rar K \ell^+ \ell^-$}
\def\tepm{$B \rar K \mu^+ \mu^-$}
\def\tept{$B \rar K \tau^+ \tau^-$}
\def\ds{\displaystyle}

\title{ {\small {\bf 
{\Large $B_c \rar \ell \bar \nu \gamma$}  DECAY IN LIGHT CONE QCD } } }

\author{ {\small T. M. AL\.{I}EV$^1$ \,,
M. SAVCI$^2$ \thanks
{e-mail: savci@rorqual.cc.metu.edu.tr} }\\
{\small 1) Physics Department, Girne American University} \\
{\small Mersin--10, Turkey }\\
{\small 2) Physics Department, Middle East Technical University} \\
{\small 06531 Ankara, Turkey} }

\date{}

\begin{titlepage}
\maketitle
\thispagestyle{empty}

\begin{abstract}
\baselineskip  0.7cm
The radiative $B_c \rar \ell \bar \nu \gamma$ decay is investigated in the 
Standard Model in framework of the light cone QCD. The transition form
factors and decay width are calculated. A comparison of light cone QCD and
constituent quark model predictions on these quantities is presented. 

\end{abstract}

\vspace{1cm}
\end{titlepage}

\section{Introduction}
The experimental and theoretical investigations of the heavy flavored
hadrons constitutes one of the main research area in particle physics. This
is due their outstanding role in the precise determination of the
fundamental parameters of the Standard Model (SM), such as
Cabibbo--Kobayashi--Maskawa (CKM) matrix elements, leptonic decay constants
etc., and for deeper understanding of the dynamics of QCD. In this sense,
$B_c$ mesons occupy an exceptional place. Since $B_c$ mesons contain two heavy
quarks, their decay channels are very rich compared to that of 
$B_q,~(q=u,~d,~s)$ mesons.
Moreover, QCD predictions on $B_c$ meson decays are more reliable since for
the heavy quarks perturbation theory works quite well. Because of the
above--mentioned reasons, the investigation of the properties of $B_c$
mesons receives special attention. Production and different decay channels
of $B_c$ mesons are widely discussed in the current literature (for a
review, see \cite{R1}). In \cite{R2,R3}, the number of $B_c$ mesons that 
will be produced in LHC, is estimated to be $\sim 2 \times 10^8$. This is
clearly an indication of the real possibility of an experimental investigation 
of the properties of the $B_c$ mesons at LHC. 
In addition to that, the $B_c$ meson decay channels can bring about some
background contribution to the $B^\pm$ meson decays, with the same final
states \cite{R4}, which is another reason that makes the
precise study of the decay channels of $B_c$ mesons worthwhile.

The pure leptonic decays of $B_c$ mesons are the simplest among all decays.
In principle the pure leptonic decay $B_c^- \rar \ell \bar \nu$ can be used
for the determination of the leptonic decay constant $f_{B_c}$, which is 
one of the fundamental parameters of the hadron physics. But this type of
decays are helicity suppressed by a factor of $m_\ell^2/m_{B_c}^2$, and hence
a precise determination of $f_{B_c}$ is difficult to make. On the other
hand, although the $B_c^- \rar \tau \bar \nu$ channel is free of the
above--mentioned helicity suppression, its observation is possible if we
have good efficiency. 

When a photon is radiated in addition to the leptons, the helicity
suppression is removed and a large branching ratio is expected, which is the
reason why it makes the investigation of the 
$B_c^- \rar \ell \bar \nu \gamma$ decay much more interesting.

Note that, the $B_c^- \rar \ell \bar \nu \gamma$ decay is investigated
in the SM within the context of the constituent quark model approach
\cite{R5}. But the concept of the "constituent quark mass" is itself poorly
understood and its relation with QCD is unclear. Therefore the prediction of the
branching ratio within the context of the constituent quark model approach
is strongly model dependent. Our aim in  this work is to investigate the 
$B_c^- \rar \ell \bar \nu \gamma$ decay in a model independent way, namely
within the framework of the light cone QCD sum rules (for a review on light
cone QCD sum rules, see \cite{R6}).

The paper is organized as follows. In Section 2 we derive the sum rules for
the transition form factors, which appears in 
$B_c^- \rar \ell \bar \nu \gamma$ decay. In Section 3 we present the
numerical analysis and give a brief discussion about the results.

\section{Light cone QCD sum rules for transition form factors}

The $B_c^- \rar \ell \bar \nu$ decay at quark level is described by the tree
level Feynman diagram, which is presented in Fig. 1. The corresponding
effective Hamiltonian for the $B_c^- \rar \ell \bar \nu$ decay is
\bea
{\cal H} = \frac{G}{\sqrt{2}} V_{cb} \bar c \gamma_\mu 
\ga 1 - \gamma_5 \dr b \bar \ell \gamma_\mu \ga 1 - \gamma_5 \dr \nu~.
\eea

As already noted, the pure leptonic process 
$B_c^- \rar \ell \bar \nu~ \ga \ell = e,~\mu \dr$ is helicity suppressed. If
a photon is attached to any of the charged lines, no helicity suppression
exists. However if a photon is attached to the charged lepton line, it
follows from helicity arguments that the contribution of such a diagram must
be proportional to the lepton mass $m_\ell$, and hence it can safely be
neglected. When a photon is radiated from $W$--boson line, the contribution
of this diagram is also strongly suppressed by a factor $m_b^2/m_W^2$, since
an extra $1/m_W^2$ factor comes from the second $W$--boson propagator.
Therefore in $B_c^- \rar \ell \bar \nu \gamma$, the main 
contribution should come from the diagrams, where photon is radiated from the
initial quark lines. Thus the corresponding matrix element for the 
$B_c^- \rar \ell \bar \nu \gamma$ process can be written as
\bea
{\cal M} = \la \gamma(q) \ve \bar c \gamma_\mu \ga 1-\gamma_5 \dr b \ve
B_c(p+q) \ra \bar \ell \gamma_\mu \ga 1-\gamma_5 \dr \nu ~.
\eea
The matrix element 
$\left<\gamma(q) \vel \bar c \gamma_\mu 
\ga 1 - \gamma_5 \dr \ver B(p+q) \right>$ describes the annihilation of the 
$B_c$ mesons into the current $\bar c \gamma_\mu \ga 1-\gamma_5 \dr b$, with
momentum $p$ accompanied by the radiation of a real photon with momentum
$q$.
This matrix element can be written in terms of two independent, 
gauge invariant structures:
\bea
\la \gamma(q) \ve \bar c \gamma_\mu \ga 1-\gamma_5 \dr b \ve
B_c(p+q) \ra = 
e\, \Bigg\{ \epsilon_{\mu\alpha\beta\sigma} e^{*\alpha} p^\beta q^\sigma 
\frac{g(p^2)}{m_{B_c}^2} +
i \, \left[e_\mu^* \ga p q \dr - \ga e^* p \dr q_\mu \right]
\frac{f(p^2)}{m_{B_c}^2} \Bigg\}~,
\eea
where $e_\mu$ and $q_\mu$ are the polarization and the four--momentum
of the photon, respectively, $p$ is the momentum transfer, $g(p^2)$ 
and $f(p^2)$ are the parity conserving and parity violating form factors.  
The main problem is to calculate the form factors $g(p^2)$ and $f(p^2)$
including their momentum dependence, and we will calculate
these form factors in the framework of light cone $QCD$ sum rules method.

We start with the following correlator function
\bea
T_\mu(p,q) = i \, \int d^4 x e^{ipx} \left< \gamma (q) \vel 
{\cal T} \left\{ \bar c (x) \gamma_\mu \ga 1- \gamma_5 \dr b(x) \, 
\bar b (0) i \gamma_5 c(0) \right\} \ver 0 \right>~.
\eea

We will present the calculation of this correlator in two different ways.
In the first approach we sandwich $T_\mu(p,q)$ in between the hadronic
states with $B_c$ meson quantum numbers, and hence we get
\bea
T_\mu(p,q) &=& e \, \frac{m_{B_c}^2 f_{B_c}}{m_b + m_c}\, 
\frac{1}{\left[ m_{B_c}^2 - \ga p + q \dr^2 \right]} \nnb \\ 
&\times& 
\Bigg\{ \epsilon_{\mu\alpha\beta\sigma}e^{*\alpha} p^\beta q^\sigma  
\frac{g(p^2)}{m_{B_c}^2} +
i \, \left[ e_\mu^* \ga p q \dr - \ga e^* p \dr q_\mu \right]
\frac{f(p^2)}{m_{B_c}^2} \Bigg\}~,
\eea
where we have used 
\bea
\la B_c \ve \bar b \, i \gamma_5 c \ve 0 \ra = 
\frac{m_{B_c}^2 f_{B_c}}{m_b + m_c}~. \nnb
\eea

The second alternative approach, on the other hand, is to calculate the 
correlator function (4) at large Euclidean momentum, where $p^2$ and 
$(p+q)^2$ are both large and negative. The Lorentz decomposition of the
correlator is 
\bea
T_\mu(p,q) = \epsilon_{\mu\alpha\beta\sigma}e^{*\alpha} p^\beta q^\sigma \, T_1
+ i \, \left[e_\mu^* \ga p q \dr - \ga e^* p \dr q_\mu \right] T_2~.
\eea
The $B_c$ meson contains two quarks which interact with the photon only
perturbatively. This point is essentially different in the 
$B^\pm \rar \ell \bar \nu \gamma$ decay in which the photon interacts with
the quarks both perturbatively and non--perturbatively (see for example
\cite{R7}). It is our aim now to
calculate these perturbative contributions. For the invariant structures 
$T_1$ and $T_2$, we write the dispersion relation in variable $(p+q)^2$ at
fixed $p^2$.
\bea
T^{(1,2)} = \int ds \, \frac{\rho^{(1,2)}(s,p^2)}
{s - (p+q)^2} + \mbox{\rm subs.
terms}~.
\eea
Here the superscript 1 and 2 corresponds to $T_1$ and $T_2$ respectively.
$\rho^{(1,2)}$ are the spectral densities and they are calculated using the
method given in \cite{R8} (for applications of this method see for example
\cite{R7,R8,R9,R10}). After lengthy calculations we get for the
spectral densities
\bea
\lefteqn{
\rho^{(1)} = e \, \frac{N_c}{4 \pi^2}\, \frac{1}{\ga s-p^2 \dr} \,  
\Bigg\{Q_b \left[ m_b \ga \mbox{ln}
\frac{1+\alpha-\beta+\lambda}{1+\alpha-\beta-\lambda} -\lambda  \dr 
+ m_c \lambda \right] } \nnb \\
&&+ Q_c \left[ m_c \ga \mbox{ln}
\frac{1-\alpha+\beta+\lambda}{1-\alpha+\beta-\lambda}-\lambda \dr
+ m_b  \lambda  \right] \Bigg\} ~, \\ \nnb \\ \nnb \\
\lefteqn{
\rho^{(2)} = e \, \frac{N_c}{4 \pi^2} \, \frac{1}{\ga s-p^2 \dr^2}\,
\Bigg\{
m_b  Q_b \Bigg[ \ga 2 m_b^2 + p^2 - s \dr \mbox{ln}
\frac{1+\alpha-\beta+\lambda}{1+\alpha-\beta-\lambda} } \nnb \\
&&-\lambda \ga 2 m_b^2 - 2 m_c^2 + p^2 (2 -\alpha +\beta) -s \dr \Bigg]  \nnb \\ 
&&+ m_c  Q_b \left[ - 2 m_b^2 \, \mbox{ln}
\frac{1+\alpha-\beta+\lambda}{1+\alpha-\beta-\lambda}+
\lambda \ga 2 m_b^2 - 2 m_c^2 - p^2 (\alpha -\beta) + s \dr \right]  \\
&&+ m_b  Q_c \left[ 2 m_c^2 \, \mbox{ln}
\frac{1-\alpha+\beta+\lambda}{1-\alpha+\beta-\lambda}+
\lambda \ga 2 m_b^2 - 2 m_c^2 - p^2 (\alpha -\beta) - s \dr \right] \nnb \\
&&+ m_c  Q_c \left[ \ga s - p^2 - 2 m_c^2 \dr \mbox{ln}
\frac{1-\alpha+\beta+\lambda}{1-\alpha+\beta-\lambda}-
\lambda \ga 2 m_b^2 - 2 m_c^2 - p^2 (2 + \alpha -\beta) + s \dr \right] 
\Bigg\} ~,\nnb \\ \nnb
\eea
where $ \lambda = \sqrt{1 +\alpha^2 + \beta^2 - 2 \alpha - 2 \beta 
-2 \alpha \beta}$ and $\alpha=m_b^2/s$, $\beta=m_c^2/s$. $Q_b$ and $Q_c$ are
the electric charges of the $b$ and $c$ quarks, respectively and $N_c$ is
the color factor and $\delta^\prime(s-t) = \frac{d}{dt} \delta(s-t)$. 
Using Eqs. (7), (8) and (9), $T_1$ and $T_2$ take the
following form
\newpage
\bea
\lefteqn{
T_1 = e \, \frac{N_c}{4 \pi^2} \int \frac{ds}{\left[s-\ga p+q \dr^2\right]
\left[s-p^2\right]} }  \\
&&\times
\Bigg\{Q_b \left[ m_b \ga \mbox{ln}
\frac{1+\alpha-\beta+\lambda}{1+\alpha-\beta-\lambda} -\lambda  \dr 
+ m_c \lambda \right]
+ Q_c \left[ m_c \ga \mbox{ln}
\frac{1-\alpha+\beta+\lambda}{1-\alpha+\beta-\lambda}-\lambda \dr
+ m_b  \lambda  \right] \Bigg\}~,\nnb \\ \nnb \\ \nnb \\ \nnb
\lefteqn{
T_2 = e \, \frac{N_c}{4 \pi^2} \int \frac{ds}{\left[ s-\ga p+q \dr^2 \right]
\left[ s-p^2 \right]^2}} \nnb \\
&&\times \Bigg\{
m_b  Q_b \left[ \ga 2 m_b^2 + p^2 - s \dr \mbox{ln}
\frac{1+\alpha-\beta+\lambda}{1+\alpha-\beta-\lambda} -
\lambda \ga 2 m_b^2 - 2 m_c^2 + p^2 (2 -\alpha +\beta) -s \dr \right] \nnb \\ 
&&+ m_c  Q_b \left[ - 2 m_b^2 \, \mbox{ln}
\frac{1+\alpha-\beta+\lambda}{1+\alpha-\beta-\lambda}+
\lambda \ga 2 m_b^2 - 2 m_c^2 - p^2 (\alpha -\beta) + s \dr \right] \\
&&+ m_b  Q_c \left[ 2 m_c^2 \, \mbox{ln}
\frac{1-\alpha+\beta+\lambda}{1-\alpha+\beta-\lambda}+
\lambda \ga 2 m_b^2 - 2 m_c^2 - p^2 (\alpha -\beta) - s \dr \right] \nnb \\
&&+ m_c  Q_c \left[ \ga s - p^2 - 2 m_c^2 \dr \mbox{ln}
\frac{1-\alpha+\beta+\lambda}{1-\alpha+\beta-\lambda}-
\lambda \ga 2 m_b^2 - 2 m_c^2 - p^2 (2 + \alpha -\beta) + s \dr \right] 
\Bigg\} ~.~~~\nnb
\eea
To match the results (10) and (11) with the $B_c$ meson contribution to the
correlator function (5), we introduce a new variable 
$u = \left[ \ga m_b + m_c \dr^2 - p^2\right] / \ga s-p^2 \dr$ in the dispersion
integral and invoke the quark--hadron duality in order to subtract continuum
contribution, which modeled as perturbation contribution starting from some
threshold $s_0$ and finally perform the Borel transformation in the variable 
$(p + q)^2$. This operation is necessary to suppress higher states and
continuum contributions. As a result we arrive at the following sum rules
for the form factors $g(p^2)$ and $f(p^2)$:
\bea
\lefteqn{
g(p^2) = \frac{m_b+m_c}{f_{B_c}} \, \frac{N_c}{4 \pi^2} \int_\Delta^1 
\frac{du}{u} \, e^{\left[m_{B_c}^2 u  - \ga m_b+m_c \dr^2 + 
p^2 \bar u \right]/ \ga M^2 u \dr} } \\
&&\times \Bigg[ \ga Q_c - Q_b \dr \ga m_b - m_c \dr \lambda +
Q_b m_b \, \mbox{ln} \frac{1+\alpha-\beta+\lambda}{1+\alpha-\beta-\lambda} +
Q_c m_c \, \mbox{ln} \frac{1-\alpha+\beta+\lambda}{1-\alpha+\beta-\lambda}
\Bigg]~, \nnb \\ \nnb \\ \nnb \\ \nnb
\lefteqn{
f(p^2) = \frac{m_b+m_c}{f_{B_c}} \, \frac{N_c}{4 \pi^2} \int_\Delta^1
\frac{du}{\left[ \ga m_b + m_c \dr^2 - p^2 \right]} \,
e^{\left[m_{B_c}^2 u - \ga m_b+m_c \dr^2 + 
p^2 \bar u \right]/ \ga M^2 u \dr } } \nnb \\
&&\times \Bigg\{
m_b  Q_b \left[ \ga 2 m_b^2 + p^2 - s \dr \mbox{ln}
\frac{1+\alpha-\beta+\lambda}{1+\alpha-\beta-\lambda} -
\lambda \ga 2 m_b^2 - 2 m_c^2 + p^2 (2 -\alpha +\beta) -s \dr \right] \nnb \\ 
&&+ m_c  Q_b \left[ - 2 m_b^2 \, \mbox{ln}
\frac{1+\alpha-\beta+\lambda}{1+\alpha-\beta-\lambda}+
\lambda \ga 2 m_b^2 - 2 m_c^2 - p^2 (\alpha -\beta) + s \dr \right] \\
&&+ m_b  Q_c \left[ 2 m_c^2 \, \mbox{ln}
\frac{1-\alpha+\beta+\lambda}{1-\alpha+\beta-\lambda}+
\lambda \ga 2 m_b^2 - 2 m_c^2 - p^2 (\alpha -\beta) - s \dr \right] \nnb \\
&&+ m_c  Q_c \left[ \ga s - p^2 - 2 m_c^2 \dr \mbox{ln}
\frac{1-\alpha+\beta+\lambda}{1-\alpha+\beta-\lambda}-
\lambda \ga 2 m_b^2 - 2 m_c^2 - p^2 (2 + \alpha -\beta) + s \dr \right] 
\Bigg\} ~,\nnb
\eea   
where in Eqs. (12) and (13) 
\bea
s = \frac{\ga m_b+m_c \dr^2 - p^2 \bar u}{u}~,~~~
\bar u = 1-u~,~~~\mbox{and}~~~
\Delta = \frac{\ga m_b+m_c \dr^2 - p^2}{s_0 - p^2}~, \nnb
\eea
and $s_0$ is the continuum threshold.
If we formally set $m_c \rar 0$ in Eqs. (12) and (13), the resulting 
expressions are expected to coincide with the form factors 
corresponding only to the perturbative part of the  
$B^\pm \rar \ell \bar \nu \gamma$ decay. 
This decay was investigated in \cite{R7}, and their results are identical to
ours in the $m_c \rar 0$ limit. 

The calculation for the differential decay rate yields
\bea
\frac{d\Gamma}{dx} = \frac{G^2 \alpha}{96 \pi^2} \ve V_{bc} \ve^2 m_{B_c}^3
x \ga 1-x \dr^3  
\left[ \ve g(x) \ve^2 + \ve f(x) \ve^2 \right] ~,
\eea
where $x = p^2/m_{B_c}^2$.
The corresponding decay width which follows from the above expression is
\bea
\Gamma = \frac{G^2 \alpha}{96 \pi^2} \ve V_{bc} \ve^2 m_{B_c}^3
\int_0^1 x \ga 1-x \dr^3 \left[\ve g(x) \ve^2 + \ve f(x) \ve^2 \right] dx ~.
\eea

\section{Numerical analysis}
In regard to the numerical analysis in the evaluation of the form factors,
we have used the following set of parameters:
$m_b=4.7~GeV,~m_c=1.4~GeV,~m_{B_c}=6.258~GeV$ \cite{R1,R11,R12},
$s_0=50~GeV^2$ and $f_{B_c}=0.35~GeV$ \cite{R11,R12,R13}.

In Fig. 2 we present the dependence of $g(p^2=0)$ and $f(p^2=0)$ on the
Borel parameter $M^2$. From this figure we see that the best stability
is achieved in the range $15~GeV^2 < M^2 < 20~GeV^2$ for which an
uncertainty less than $10\%$ is observed. The analysis for $p^2 \ne 0$
yields similar results as well.

The sum rules of the type (12) and (13) are expected to work in the region
$\ga m_b + m_c \dr^2 - p^2 \sim$ few $GeV^2$, which is smaller than the
maximal available $p^2=\ga m_b + m_c \dr^2$. To extend our results to the
whole region of $p^2$ some extrapolation has been used. The best fits are
achieved with the following pole formulas:
\bea
g\ga p^2 \dr = \frac{g(0)}{1 - p^2/m_1^2}~,~~~~~
f\ga p^2 \dr = \frac{f(0)}{1 - p^2/m_2^2}~, 
\eea
where
\bea
g(0) &=& 0.44 \pm 0.04~GeV,~~~~m_1^2 = 43.1~GeV^2~, \nnb \\
f(0) &=& 0.21 \pm 0.02~GeV,~~~~m_2^2 = 48.0~GeV^2~.
\eea
For a more precise determination of the form factors, the perturbative
$ O ( \alpha_s ) $ corrections need to be calculated. We will consider this
point elsewhere in a future work.

Fig. 3 depicts the dependence of the differential branching ratio on $x$.
It is observed from this figure that the photon spectra 
$\ga x = 1 - 2 E_\gamma /m_{B_c} \dr$ is practically symmetric. Note that
the quark model approach also predicts a symmetric spectra for the photon
(see Fig. 2 in \cite{R5}).

Finally, we summarize the numerical results of the branching ratios. Taking
$\ve V_{cb} \ve = 0.04$ \cite{R14}, $\tau \ga B_c \dr = 0.52 \times
10^{-12}~s$ \cite{R15}, we get 
\bea
{\cal B} \ga B_c \rar \ell \bar \nu \gamma \dr = 1.0 \times 10^{-5}~.
\eea
Here we note that the quark model approach yields \cite{R5} 
\bea
{\cal B} \ga B_c \rar \ell \bar \nu \gamma \dr = 5.0 \times 10^{-5}~.
\eea
If we compare Eq. (18) and (19), it is obvious that the light cone QCD sum
rules prediction on the branching ratio is approximately five (three)
times smaller than the one predicted by the quark model approach, 
if the constituent $u$--quark mass is taken as $0.35~GeV$ ($0.48~GeV$).  

For completeness we present the predictions of the branching ratios for the
pure leptonic decays \cite{R5}:
\bea
{\cal B} \ga B_c \rar \mu \bar \nu_\mu \dr &=& 6.0 \times 10^{-5}~, \nnb \\
{\cal B} \ga B_c \rar e \bar \nu_e \dr &=& 1.4 \times 10^{-9}~.
\eea
From a comparison of Eqs. (18) and (20), we observe that the branching ratio of
the radiative leptonic decay is of same order with the corresponding pure 
leptonic decay (i.e., for $\mu$ case). 

Since $B_c \rar \ell \bar \nu \gamma$ and $B^\pm \rar \ell \bar \nu \gamma$
decays have the same final states, it will be interesting to know about the
relative fraction of the $\ell \bar \nu \gamma$ final states which outcomes from
the different sources of $B_c$ and $B^\pm$.

a) QCD sum rules prediction:
\bea
\frac{N_{B^\pm}}{N_{B_c}} \simeq 0.2~, \nnb
\eea
where, for the branching ratio of the  $B^\pm  \rar \ell \bar \nu \gamma $ 
decay we have used ${\cal B} \ga B^\pm  \rar \ell \bar \nu \gamma \dr
\simeq 2.0 \times 10^{-6}$ \cite{R7}.

b) The quark model prediction (see \cite{R5}):
\bea
\frac{N_{B_u}}{N_{B_c}} \simeq  
\left\{ \begin{array}{ll} 
~~ 1.2 & \ga \mbox{for}~ m_u = 0.35~GeV \dr~, \\ \\
~~ 0.7 & \ga \mbox{for}~ m_u = 0.48~GeV \dr~. 
\end{array} \right.
\eea   

It is quite obvious from the above results that, the two approaches lead to
absolutely different predictions on the relative fraction of the $\ell \bar
\nu \gamma$ final states from $B_c$ and $B^\pm$. This result will be checked
in LHC in future experiments. According to the estimations that have been
made [2--3], the number of $B_c$ mesons that are expected to be produced 
in LHC is $\sim 2 \times 10^8$, and hence the $B_c \rar \ell \bar \nu \gamma$  
decay can easily be detected at LHC which opens the way for a real possibility 
of the experimental investigation of the properties of the $B_c$ meson.
In an investigation of the $B_c \rar \ell \bar \nu \gamma$ decay, according
to our estimation, the background contribution from 
$B^\pm  \rar \ell \bar \nu \gamma $ is small.

In conclusion, we have calculated the branching ratio for the 
$B_c \rar \ell \bar \nu \gamma$ decay, in the framework of the QCD sum rules
method within SM and found that
${\cal B}\ga B_c \rar \ell \bar \nu \gamma \dr \simeq 1.0 \times 10^{-5}$.

\newpage

\section*{Figure Captions}
{\bf 1.} Feynman diagram for the $B_c \rar \ell \bar \nu$. \\ \\
{\bf 2.} Dependence of the form factors $g \ga p^2=0 \dr$ and 
$f \ga p^2=0 \dr$ on the Borel parameter $M^2$. In this graph, the continuum
threshold $s_0$ is fixed to the value of $50~GeV^2$.\\ \\
{\bf 3.} Dependence of the differential Branching ratio on $x$ for the
$B_c \rar \ell \bar \nu \gamma$ decay.

\begin{figure}
\vspace{25.0cm}
    \includegraphics{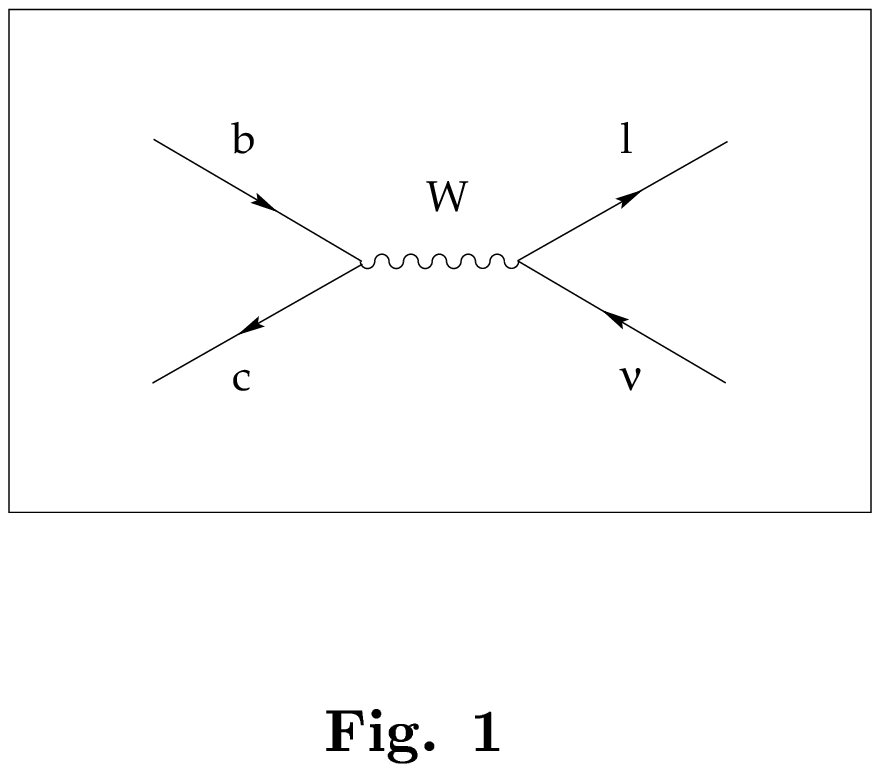}
    \includegraphics{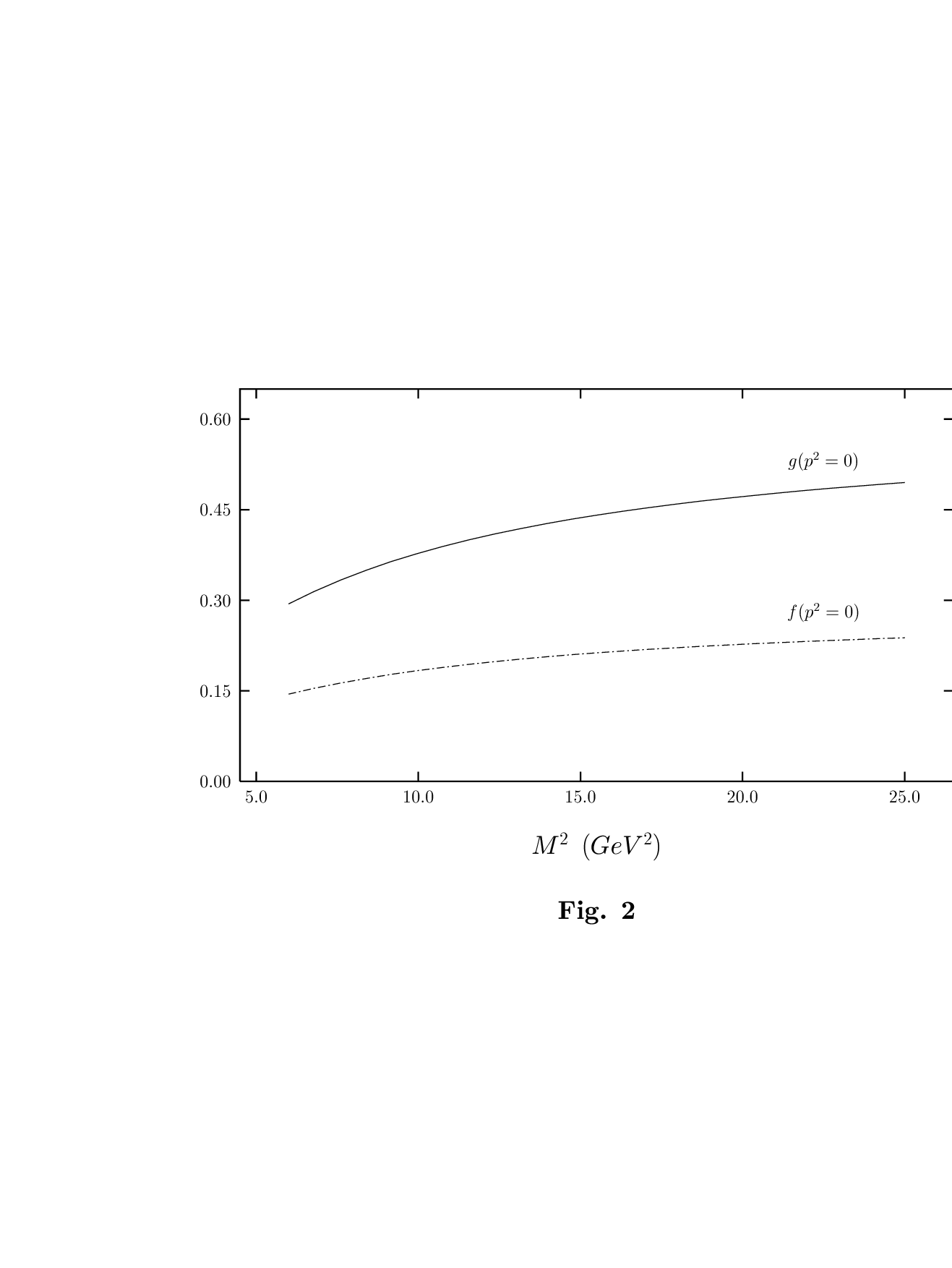}
    \vspace{-4.0cm}
\vspace{0.0cm}

\end{figure}

\begin{figure}
\vspace{25.0cm}
    \includegraphics{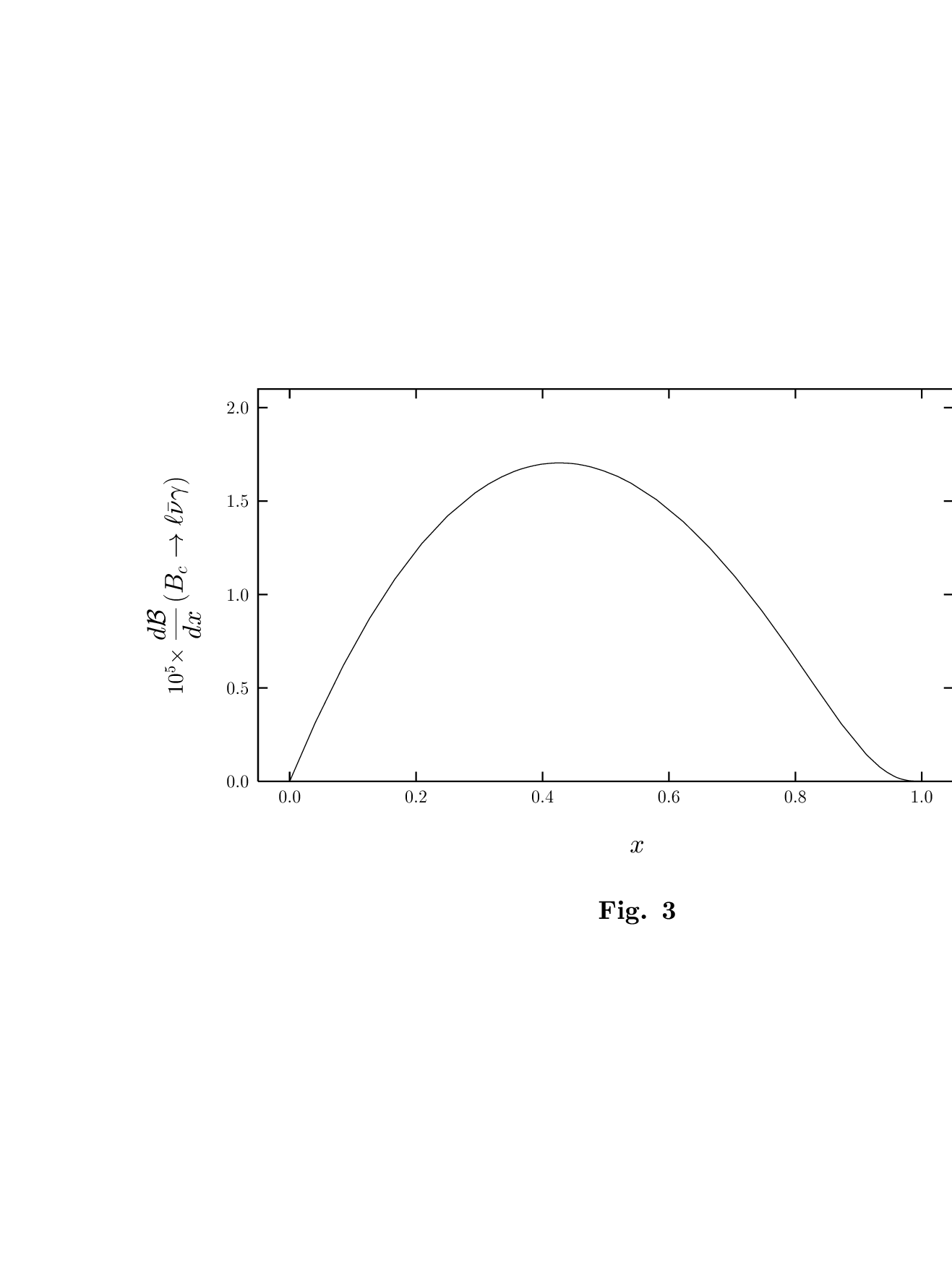}
    \vspace{-4.0cm}
\vspace{0.0cm}

\end{figure}


\newpage

\begin{thebibliography}{99}

\bibitem{R1} S. S. Gershtein, V. V. Kiselev, A. K. Likhoded, A. V. Tkabladze,\\
{\it Phys. Usp.} {\bf 38} (1995) 1.

\bibitem{R2} Chao--Hsi Chang, Yu-Qi Chen, 
{\it Phys. Rev.} {\bf D48} (1993) 4086.

\bibitem{R3} K. Cheung, 
{\it Phys. Rev. Lett.} {\bf 71} (1993) 3413.

\bibitem{R4} Michelangelo L. Mangano, S. R. Slabospitskii,
{\it preprint} {\bf hep-ph}/9707248.

\bibitem{R5} Chao--Hsi Chang, Jian-Ping Cheng Cai-Dian L\"{u},
{\it preprint} {\bf hep-ph}/9712325.

\bibitem{R6} V. M. Braun, 
{\it preprint} {\bf hep-ph}/9801222.

\bibitem{R7} G. Eilam, I. Halperin, R. R. Mendel,
{\it Phys. Lett.} {\bf B361} (1995) 137.

\bibitem{R8} V. A. Nesterenko, A. V. Radyushkin,
{\it Sov. J. Nucl. Phys.} {\bf 39} (1984) 811.

\bibitem{R9} T. M. Aliev, D. A. Demir, E. \.{I}ltan and N. K. Pak,\\
{\it Phys. Rev.} {\bf D54} (1996) 857.

\bibitem{R10} V. M. Belyaev, V. M. Braun, A. Khodjamirian, R. R\"{u}ckl,\\
{\it Phys. Rev.} {\bf D51} (1995) 6177.

\bibitem{R11} V. V. Kiselev, A. V. Tkabladze,
{\it Phys.Rev.} {\bf D48} (1993) 5208.

\bibitem{R12} P. Colangelo, G. Nardulli, N. Paver,
{\it Z. Phys.} {\bf C57} (1993) 43.

\bibitem{R13} T. M. Aliev, O. Y{\i}lmaz,
{\it Nuovo Cimento} {\bf 105A} (1992) 827.

\bibitem{R14}  Particle Data Group,
{\it Phys. Rev.} {\bf D54} Part II (1996).

\bibitem{R15} M. Beneke, G. Buchalla,
{\it Phys. Rev.} {\bf D53} (1996) 4991.
 
\end{thebibliography}
\end{document}